\documentclass[12pt]{article}
\usepackage[margin=3cm]{geometry}
\usepackage{graphicx}
\usepackage{xcolor}
\usepackage{amsmath, amssymb}
\usepackage{url, hyperref}
\usepackage[font=small]{caption}

\title{\bf Bubble wall velocities with an\\ extended fluid \emph{Ansatz}}
\author{Gl\'auber C. Dorsch\thanks{glauber@fisica.ufmg.br}~\ and Daniel A. Pinto\thanks{dwavea@ufmg.br}\\[3mm]
\small \it Universidade Federal de Minas Gerais, 31270-901, Belo Horizonte, MG, Brazil}
\date{}

\begin{document}

\maketitle

\begin{abstract}
    We compute the terminal bubble wall velocity during a cosmological phase transition by modelling non-equilibrium effects in the plasma with the so-called ``extended fluid \emph{Ansatz}''. A $\phi^6$ operator is included in the Standard Model effective potential to mimic effects of new physics. Hydrodynamical heating of the plasma ahead of the bubble is taken into account. We find that the inclusion of higher order terms in the fluid \emph{Ansatz} is typically relevant, and may even turn detonation solutions into deflagrations. Our results also corroborate recent findings in the literature that, for a Standard Model particle content in the plasma, only deflagration solutions are viable. However, we also show that this outcome may be altered in a theory with a different particle content.
\end{abstract}

\section{Introduction}

The detection of gravitational waves (GWs) by LIGO/Virgo/KAGRA~\cite{LIGOScientific:2016aoc, LIGOScientific:2018mvr, LIGOScientific:2021usb, LIGOScientific:2021djp}, and the strong evidence for a stochastic GW background seen by NANOGrav~\cite{NANOGrav:2023gor}, prove that we are capable of 
extracting information from messengers previously inaccessible to us,  allowing us to reconstruct events that took place a long time ago and in galaxies far away. 
Among the possible sources of these GWs, first order cosmological phase transitions play a prominent role from the perspective of particle physics. 
Because the early Universe has undergone periods of fairly high temperatures, any remnant from such ancient times would contain information on the effective particle content and on how they interact in this highly energetic regime. So a detection of cosmological GWs could place constraints on particle physics models, as long as one has an accurate prediction on how the spectrum depends on the microphysics. 
Importantly, the information carried by these novel informants could even be complementary to those obtainable from collider experiments~\cite{Caprini:2019egz, Arcadi:2023lwc}. Moreover, a number of other relics could have been produced during such a cosmological phase transition, such as a matter-antimatter asymmetry~\cite{Konstandin:2013caa, Morrissey:2012db} and a dark matter abundance~\cite{Azatov:2021ifm, Baldes:2022oev, Jiang:2023nkj, Chun:2023ezg}.

A first order phase transition takes place when bubbles of a stable phase nucleate in the plasma filled with the metastable state, then expand and percolate. Baryogenesis and dark matter could be produced during the bubble expansion, whereas at the end of the transition the mutual collisions of the bubbles break their spherical symmetry, leading to a time-dependent quadrupole moment in energy-momentum and thus generating GWs~\cite{Caprini:2018mtu, Hindmarsh:2020hop}. A crucial parameter for determining the density of relics produced is the velocity of the bubble expansion, which depends on non-equilibrium dynamics taking place during the phase transition. The passage of the bubble drives the plasma out of equilibrium, whose backreaction acts as an effective friction against the expansion. Determining this friction term is crucial for computing the terminal wall velocity. 

A number of important advances have been made in the recent literature regarding the calculation of the wall velocity. In the limit of ultrarelativistic runaway walls, all particles in the plasma hit the bubble with enough energy to overcome the barrier. There are no particles reflected back to the symmetric phase, and no information about the arrival of the bubble has reached the plasma ahead of the wall yet, so equilibrium considerations suffice to determine the backreaction against the bubble~\cite{Bodeker:2017cim, Bodeker:2009qy, Hoche:2020ysm,Azatov:2020ufh,Cai:2020djd, Gouttenoire:2021kjv, Ai:2023suz, Azatov:2023xem}. On the other hand, non-equilibrium effects may be relevant in the non-relativistic regime. In this case, one has to solve the corresponding integro-differential Boltzmann equation for the distribution functions to find the corresponding pressure difference across the bubble wall. To make this task feasible, a common approach is to make an Ansatz for the shape of the non-equilibrium distribution function, performing a series expansion and truncating it at some appropriate order to achieve closure of the system of equations~\cite{Moore:1995si, Cline:2020jre, Laurent:2020gpg, Dorsch:2021nje, Dorsch:2021ubz}. A reasonable first approximation is to assume that the plasma behaves as a perfect fluid, with small local fluctuations in chemical potential, temperature and velocity~\cite{Moore:1995si}. This approximation has prevailed in the literature for a long time~\cite{Joyce:1994zt, Bodeker:2004ws, Fromme:2006wx, Huber:2011aa, Huber:2013kj, Kozaczuk:2015owa, Dorsch:2016nrg, Dorsch:2018pat}, until it was pointed out that it gives inconsistent results for baryogenesis calculations~\cite{Cline:2020jre}. 

A so-called ``new formalism'' to modelling non-equilibrium effects was then presented in~\cite{Cline:2020jre} and applied to the calculation of the friction and the wall velocity~\cite{Laurent:2020gpg, Lewicki:2021pgr}. However, it was later shown that this ``new formalism'' is also inconsistent~\cite{Dorsch:2021nje}, since it assumes the perfect fluid Ansatz for computing the collision integrals but postulates different momentum dependencies in the other terms of the Boltzmann equation. A fully consistent approach to the computation of friction, extending the fluid Ansatz beyond the perfect fluid approximation (i.e. including dissipative effects), has been performed in~\cite{Dorsch:2021ubz}. However, studies on the behaviour of the wall velocity in this context are still lacking\footnote{There have been other attempts present in the recent literature, such as an Ansatz expanding in Chebyshev polynomials~\cite{Laurent:2022jrs}, as well as fully numerical methods~\cite{DeCurtis:2022hlx, DeCurtis:2023hil}.}.

In light of these recent developments, it is the purpose of this paper to investigate the behaviour of the wall velocity in the presence of dissipative effects in the fluid Ansatz. For simplicity, we will consider a one-field model with an effective $\phi^6$ operator, mimicking effects from Standard Model extensions (see e.g.~\cite{Jiang:2018pbd}). Our findings indicate that the inclusion of higher order terms in the fluid Ansatz is relevant, and could even turn a detonation into a deflagration. For a Standard Model particle content in the plasma, our findings corroborate other results of the literature that, due to the hydrodynamical heating of the plasma ahead of the wall, the only viable solutions to the bubble expansion would correspond to deflagrations or luminal detonations, i.e. with $v_w=1$~\cite{Cline:2021iff, Laurent:2022jrs, Krajewski:2023clt}. This effect has been known in the literature for some time under the name of hydrodynamical obstruction~\cite{Konstandin:2010dm}. However, we will see that this result is heavily dependent on the particle content of the plasma.

The paper is organized as follows. In section~\ref{sec:friction} we present the model and review the formalism that leads to the Klein-Gordon equation in a thermal plasma at non-equilibrium. This equation includes so-called friction terms stemming from the disturbance of the plasma from equilibrium due to the passage of the bubble. In section~\ref{sec:Boltzmann} we review the Boltzmann approach to computing these non-equilibrium distribution functions in the so-called ``extended fluid \emph{Ansatz}'' introduced recently in~\cite{Dorsch:2021nje, Dorsch:2021ubz} (see also~\cite{DeGroot:1980dk}). Section~\ref{sec:hydro} focuses on the macroscopic hydrodynamical effects and the calculation of the temperature immediately ahead of the bubble wall, which is where the effects that lead to friction take place. We present our results in section~\ref{sec:results} and conclude in section~\ref{sec:conclusions}.

\section{Friction and non-equilibrium dynamics}
\label{sec:friction}

We will consider a cosmological phase transition described by the dynamics of a single scalar field immersed in a plasma of Standard Model particles. We will take the scalar potential at zero temperature to be
\begin{equation}
    V_0(\phi) = -\mu^2 \frac{\phi^2}{2} + \lambda \frac{\phi^4}{4}  + \frac{\phi^6}{8M^2} + \left(\sum_{i} \frac{g_i}{64\pi^2} \frac{m_i^4}{v^4}\right) \phi^4\log\frac{\phi^2}{v^2},
\end{equation} 
which describes self-interactions of the scalar field plus 1-loop corrections, i.e. a Coleman-Weinberg term. The $\phi^6$ operator encapsulates effects from unknown physics at an energy scale $M$, which we take as a free parameter of the model. The other parameters, $\mu^2$ and $\lambda$, are fixed so that this potential has a minimum at $v=246.22$~GeV and the scalar field mass is $m_h=125$~GeV, yielding
\begin{equation}
    \mu^2 = \frac{m_h^2}{2} - \dfrac{3v^4}{4M^2}
    \quad\text{and}\quad
    \lambda = \dfrac{m_h^2}{2v^2} - \dfrac{3v^2}{4M^2}.
\end{equation}
We will do a two-fold analysis to evaluate the impact of the particle content of the theory in the wall velocity. In one case, we will consider only tops as heavy particles running in the loop, whereas in a second situation we will consider tops, $W$'s and $Z$'s running in the loop, with $g_i = 12, 6, 3$ their respective number of degrees of freedom and masses $m_i = 173.1~\text{GeV}, 80.385~\text{GeV}$ and $91.1876$~GeV respectively.

\begin{figure}
    \centering
    \raisebox{15pt}{\includegraphics[scale=.7]{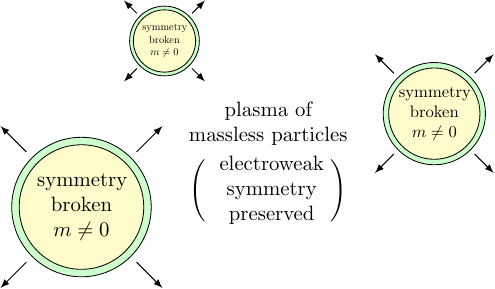}}
    \qquad\qquad
    \includegraphics[trim=0 0 0 0, clip, scale=.45]{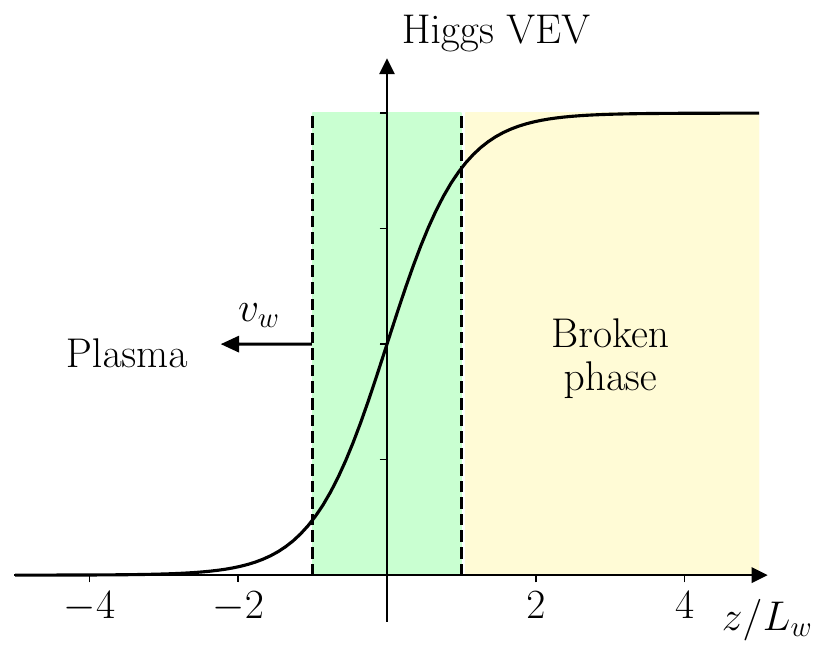}
    \caption{A first order phase transition proceeds through the nucleation of bubbles. (Left) Inside the bubbles, the scalar field acquires a VEV and particles become massive. Outside, the symmetry is still preserved, and particles remain massless. (Right) Profile of the scalar field along the bubble, which is also a depiction of the spatial dependence of particle masses. The horizontal axis is in units of the bubble wall thickness $L_w$, and the wall moves to the left with a velocity $v_w$ with respect to the plasma.}
    \label{fig:EWPTshape}
\end{figure}

Conservation of the total energy-momentum of the scalar field and of the plasma leads to the Klein-Gordon equation~\cite{Konstandin:2014zta}
\begin{equation}
    \Box \phi\,\partial_\nu \phi + \dfrac{\partial V_0}{\partial\phi}\partial_\nu \phi 
    + \sum_i \dfrac{g_i}{2} \frac{\partial m_i^2}{\partial\phi}\partial_\nu \phi \int \frac{d^3 p}{(2 \pi)^3 E_i} f_i(p,x) = 0,
    \label{eq:eom2}
\end{equation}
with $f_i(p,x)$ the distribution function of particle $i$ and $g_i$ its number of degrees of freedom. We can decompose $f_i(p,x)$ as a sum of an equilibrium and a non-equilibrium part,
\begin{equation}
    f_i(p,x) = f^{\rm eq}_i(p) + \delta f_i(p,x)
    \qquad\text{with}\qquad
    f_i^{\rm eq}(p) = \dfrac{1}{e^{\beta p^\mu u_\mu}\mp 1},
    \label{eq:fdf}
\end{equation}
with $u_\mu$ the plasma four-velocity, $\beta\equiv 1/T$ its inverse temperature, and the sign $\mp$ stands for bosons and fermions, respectively. 
In the plasma rest frame, $p^\mu u_\mu = E = \sqrt{\mathbf{p}^2+m^2}$, so
\begin{equation}\begin{split}
    \frac12\frac{\partial m_i^2}{\partial\phi} \int \frac{d^3 p}{(2 \pi)^3 E_i} f_i^{\rm eq}(p)
    &= \pm T\frac{\partial}{\partial\phi} \int \frac{d^3 p}{(2 \pi)^3} \log\bigg(1 \mp e^{-\beta\sqrt{\mathbf{p}^2 + m^2}}\bigg)
\end{split}\end{equation}
and this term can be absorbed into $V_0$ to yield the effective thermal potential
\begin{equation}
    V(T,\phi) \equiv V_0 + \sum_i (\pm g_i) \frac{T^4}{2\pi^2}\int_0^\infty dp\, p^2\,\log\bigg(1 \mp e^{-\beta\sqrt{p^2 + (m/T)^2}}\bigg),
    \label{eq:Veff}
\end{equation}
where the upper sign corresponds to bosons and the bottom one to fermions. Then the Klein-Gordon equation for the scalar field immersed in the plasma can finally be written as
\begin{equation}
    \Box \phi\,\partial_\nu \phi + \dfrac{\partial V(T,\phi)}{\partial\phi}\partial_\nu \phi 
    + \sum_i \dfrac{g_i}{2} \frac{\partial m_i^2}{\partial\phi}\partial_\nu \phi \int \frac{d^3 p}{(2 \pi)^3 E_i} \delta f_i(p, x) = 0.
    \label{eq:eom}
\end{equation}

The last term arises from interactions of the plasma with the bubble wall. As the wall passes by, the plasma particles acquire masses and their energy spectrum shifts. The equilibrium configuration is different far ahead and far behind the bubble wall, and in the region in-between the particles must obey a non-equilibrium distribution function. When the wall width $L_w$ is much larger than typical momenta of the incident particles, $p\sim T$, a WKB approximation can be applied to the processes taking place along the bubble wall, and it can be shown that the particle distribution functions satisfy the semi-classical Boltzmann equation~\cite{Konstandin:2013caa}
\begin{equation}
    \left( p^\nu \partial_\nu  + \dfrac{\partial_\nu m_i^2}{2} \partial_{p_\nu} \right) f_i + \mathcal{C}_i[f] = 0,
    \label{eq:Boltzmann}
\end{equation}
where $\mathcal{C}_i$ are collision terms affecting the distribution of particle $i$, but depending on the distribution of all other particles interacting with it\footnote{Notice that the collision terms do not appear in eq.~\eqref{eq:eom}, since they describe the transfer of energy-momentum between the plasma and the scalar field, while eq.~\eqref{eq:eom} expresses the conservation of total energy-momentum. That these collision terms indeed vanish when summing over all the particles and the scalar field can be checked explicitly, and follows from the total conservation of energy-momentum in scatterings and annihillations~\cite{Konstandin:2014zta, Dorsch:2021ubz}. This cancellation is also key for the behaviour of friction in the fluid Ansatz, as discussed thoroughly in~\cite{Dorsch:2021ubz}.}.

After nucleating and expanding, the bubbles eventually reach an approximately stationary and planar configuration, so the problem becomes essentially one-dimensional. Moreover, in the bubble wall frame, the observer will only see a steady flow of particles from the plasma into the bubble, and the situation is time-independent, so we can replace $\partial_\nu \to \partial_z$. By modelling the scalar field profile as
\begin{equation}
    \phi(z) = \dfrac{\phi_0}{2}\left( 1 + \tanh\dfrac{z}{L_w}\right)
\end{equation}
the problem is reduced to solving eq.~\eqref{eq:eom} for two variables: the wall velocity $v_w$ and the width $L_w$. For that, we take two moments of this equation and normalize by appropriate factors of temperature to obtain a dimensionless quantity, namely
\begin{equation}\begin{split}
    M_1 &\equiv \dfrac{1}{T^4}\int_{-\infty}^\infty (\text{LHS of eq.}~\eqref{eq:eom})\, dz =0,\\
    M_2 &\equiv \dfrac{1}{T^5}\int_{-\infty}^{\infty} (\text{LHS of eq.}~\eqref{eq:eom})\,(2\phi(z)-\phi_0)\,dz = 0.
    \label{eq:M1M2_full}
\end{split}\end{equation}

In principle, one would evaluate the quantities in the above equations at the nucleation temperature $T_n$, defined as the temperature at which at least one bubble has nucleated per Hubble horizon. We compute this quantity using standard techniques found in the literature~\cite{Hindmarsh:2020hop, Coleman:1977py, Linde:1981zj, Mukhanov:991646}, namely solving the sphaleron equation to find the so-called ``bounce configuration'', calculating its $3$-dimensional Euclidean action $S_E(T)$ and imposing
\begin{equation}
    \dfrac{S_E(T_n)}{T_n} \approx 140,
    \label{eq:S_T}
\end{equation}
which is roughly equivalent to the condition of one bubble per horizon if the transition takes place at the electroweak scale. This is the temperature of the plasma at which the transition actually starts.

However, as the bubble interface crosses the plasma, some particles collide against it and are reflected back, leading to a slightly higher temperature $T_+ > T_n$ immediately ahead of the bubble wall. Since this is the region where the non-equilibrium effects are prominent, we evaluate the above quantities at this temperature $T_+$. We will discuss how to compute this temperature in section~\ref{sec:hydro}. 

Equations~\eqref{eq:M1M2_full} can then be written as~\cite{Konstandin:2014zta}
\begin{equation}\begin{split}
    M_1 &\equiv \dfrac{\Delta V}{T_+^4} + f = 0,\\
    M_2 &\equiv \dfrac{2}{15 (L_w T_+)^2}\left(\frac{\phi_0}{T_+}\right)^3 + \dfrac{W}{T_+^5} + g = 0,
    \label{eq:M1M2}
\end{split}\end{equation}
where $f$ and $g$ are the terms coming from integration over the non-equilibrium contributions (terms involving $\delta f_i$ in eq.~\eqref{eq:eom}), while
\begin{equation}\begin{split}
    \begin{pmatrix} \Delta V\\ W\end{pmatrix} &\equiv \int_{-\infty}^{\infty} \dfrac{\partial V(T_+, \phi)}{\partial \phi}\partial_z \phi\, \begin{pmatrix} 1\\ \phi_0\tanh({z}/{L_w}) \end{pmatrix}dz.
\end{split}\end{equation}
Note, in particular, that $\Delta V$ is (minus) the pressure difference across the wall due to the free energy released by the transition. Then equation $M_1$ has a simple interpretation: the pressure pushing the wall forward must be counterbalanced by the friction $f$. On the other hand, the integrand in the definition of $W$ is asymmetric under parity reflections around the origin (due to the $\tanh(z/L_w)$ term), so it can be seen as an overall ``stretching'' effect that tends to change the wall width. The solution of these equations are the values of $v_w$ and $L_w$ for which these forces are all balanced out.

In section~\ref{sec:Boltzmann} we will discuss how we compute the non-equilibrium particle distributions $\delta f_i$, which enter the calculation of the $f$ and $g$ terms in eq.~\eqref{eq:M1M2}. Then in section~\ref{sec:hydro} we discuss how to compute $T_+$. 

\section{Boltzmann approach with an extended fluid \emph{Ansatz}}
\label{sec:Boltzmann}

The non-equilibrium distribution functions $\delta f_i$ are found by solving the Boltzmann equation~\eqref{eq:Boltzmann}. If we restrict ourselves to $2\to 2$ processes, each with amplitude $\mathcal{M}_{pk\to p^\prime k^\prime}$, the collision term is
\begin{equation}\begin{split}
    \mathcal{C}[f_p] = \frac{1}{2}\sum_{\rm processes} &\int
				\dfrac{d^3k\, d^3p^\prime d^3k^\prime}{(2\pi)^{9} 2E_k\, 2E_{p^\prime}\, 2E_{k^\prime}} |\mathcal{M}_{pk\to p^\prime k^\prime}|^2 (2\pi)^4 \delta^4(p\!+\! k\!-\!p^\prime\!-\!k^\prime) \mathcal{P}_{pk \to p^\prime k^\prime},
	\label{eq:full_coll}
\end{split}\end{equation}
where
\begin{equation}
	\mathcal{P}_{pk \to p^\prime k^\prime} \equiv f_p f_k (1\pm f_{p^\prime}) (1\pm f_{k^\prime}) -
							f_{p^\prime} f_{k^\prime} (1\pm f_p) (1\pm f_k)
	\label{eq:P_occupation}
\end{equation}
is the population factor which takes into account how the number density of reactants and products in the plasma affect the overall rate of the reaction\footnote{Indeed, the first term being proportional to $f_p f_k$ means that a process $pk\to p^\prime k^\prime$ is more efficient if there are more reactants present in the plasma. Likewise, the number density of outgoing particles also influences the overall rate, taken into account by the factors $(1\pm f)$, which correspond to stimulated emission (for bosons, ``$+$'' sign) or Pauli blocking (fermions, ``$-$'' sign). The second term in $\mathcal{P}$, with an overall negative sign in front, accounts for the reverse process $p^\prime k^\prime \to pk$.}. Thus, one needs to solve an integro-differential equation for the $f$'s\footnote{See~\cite{DeCurtis:2022hlx, DeCurtis:2023hil} for recent attemps at fully numerical methods.}. 

To make this task feasible, we will make a so-called \emph{fluid Ansatz}, where the distribution functions are modeled as
\begin{equation}
    f(x,p) \equiv \dfrac{1}{e^{\beta (p^\mu u_\mu - \delta_p - \delta_\text{bg})} \pm 1},
    \label{eq:fluid_delta}
\end{equation}
and the fluctuations $\delta$ are expanded in powers of momenta as \begin{equation}
    \delta(x,p) = w^{0}(x) + p^\mu w^{(1)}_\mu(x) + p^\mu p^\nu w^{(2)}_{\mu\nu}(x) + \ldots .
    \label{eq:delta}
\end{equation}
Notice that we have separated the fluctuations in two parts, $\delta_p$ and $\delta_\text{bg}$. The latter are the fluctuations of the so-called ``background'' particles, i.e. particles that are so light that are not affected by the passage of the bubble. As discussed in ref.~\cite{Dorsch:2021ubz}, they can be modelled collectively as a perfect fluid with vanishing chemical potential, sharing a common temperature and velocity fluctuations. The fluctuations $\delta_p$ are those of the heavy species under consideration, and, as per the definition above, are computed relative to the fluctuations of the background. 

It is worth emphasizing that, \emph{up to this point}, this approach is fully generic: any fluctuation $\delta f$ can be put in the form~\eqref{eq:fluid_delta} for some function $\delta(x,p)$, and any such function can be expanded in powers of momenta, which is essentially an expansion in the 4d generalized version of spherical harmonics~\cite{DeGroot:1980dk}. Because these generalized spherical harmonics form a complete set in the functional space, this expansion is expected to converge for any reasonably well-behaved function $\delta f$. 

But we will make one further approximation, assuming that the fluctuations $w^{(i)}$ are sufficiently small that one can linearize 
\begin{equation}
    \delta f = \delta(x,p) (-f^{\prime\,\text{eq}}),
    \label{eq:linear}
\end{equation}
where $f^{\prime\,\text{eq}}$ is the derivative of the equilibrium distribution function shown in eq.~\eqref{eq:fdf} with respect to $p^\mu u_\mu$. Plugging this back into eq.~\eqref{eq:Boltzmann} and neglecting terms $\mathcal{O}(w^2)\sim \mathcal{O}(w \partial m^2)$ leads to the linearized Boltzmann equation
\begin{equation}\begin{split}
	&( \partial_\mu w^{(0)} + p^\rho \partial_\mu w^{(1)}_\rho + p^\rho p^\sigma \partial_\mu w^{(2)}_{\rho\sigma}  + \cdots )\ p^\mu  (-f_{0}') +\\[1mm]
	& \frac{1}{2}\sum_{\rm processes} \int
				\dfrac{d^3k\, d^3p^\prime d^3k^\prime}{(2\pi)^{9} 2E_k\, 2E_{p^\prime}\, 2E_{k^\prime}} 
				|\mathcal{M}_{pk\to p^\prime k^\prime}|^2 (2\pi)^4 \delta^4(p+k-p^\prime-k^\prime)\times\\
				&\quad\qquad\qquad\times f_{0p} f_{0k} (1\pm f_{0p^\prime}) (1\pm f_{0k^\prime})
				\left(\delta_p + \delta_k - \delta_{p^\prime} - \delta_{k^\prime}\right)
	=  u^\mu \dfrac{\partial_\mu m^2}{2T} (-f_0^\prime).
	\label{eq:Boltzmann_linear}
\end{split}\end{equation}

In order to extract information on the various $w^{(n)}$ we take moments of the above equation, multiplying by factors of $p^\alpha p^\beta p^\gamma \ldots$, projecting along the plasma four-velocity $u^\mu$ or along the perpendicular direction\footnote{Note that, when the wall reaches a planar steady state, the problem becomes $1+1$ dimensional and there is only one perpendicular direction to $u^\mu$.} $\overline{u}^\mu$, and integrating over $d^3p/(2\pi)^3 E_p$. The linearized collision terms can be computed~\cite{Dorsch:2021nje, Dorsch:2021ubz}, effectively reducing the integro-differential equation to a system of coupled differential equations that can be written as in the bubble wall frame as
\begin{equation}
    A\cdot \partial_z (q + q_\text{bg}) + \dfrac{1}{\gamma_w}\Gamma\cdot q = v_w \dfrac{\partial_z m^2}{2T^2} S,
    \label{eq:system}
\end{equation}
with
\begin{equation}
    q = (w^{(0)}, T u^\mu w^{(1)}_{\mu}, T \bar{u}^\mu w^{(1)}_{\mu}, T^2 u^\mu u^\nu w^{(2)}_{\mu\nu}, T^2 u^\mu \bar{u}^\nu w^{(2)}_{\mu\nu}, T^2 \bar{u}^\mu \bar{u}^\nu w^{(2)}_{\mu\nu}, \ldots)^T,
\end{equation}
where factors of temperature $T$ have been absorbed into $q$ to keep the perturbations dimensionless. The matrix $A$ are kinetic coefficients, $\Gamma$ describe the collision terms, and the right-hand side is the source of the fluctuations, coming from the variation of the particle's mass due to the passage of the bubble. The $\gamma_w^{-1}$ factor in the collision term corresponds to time-dilation of the interactions, as the observer sees the particles moving towards the wall with velocity $v_w$. 

For the background species the source term vanishes by definition, as does the collision terms of background species among themselves. There remains only collision terms of the background with the heavy species, $\Gamma_\text{bg, i}$, but, due to energy-momentum conservation, these are related to collision terms $\Gamma_i$ appearing in the equation for the heavy species $i$ through $\Gamma_\text{bg,i} + N_i \Gamma_i = 0$~\cite{Dorsch:2021nje, Dorsch:2021ubz}. Then the Boltzmann equation for the background fluctuations is
\begin{equation}
    \partial_z q_\text{bg} = \dfrac{1}{\gamma_w} \sum_i g_i \,(A^{-1}_\text{bg}\cdot \Gamma_i) \cdot q_i ,
    \label{eq:bg}
\end{equation}
with $g_i$ the d.o.f. of species $i$. This can then be substituted back into eq.~\eqref{eq:system} to yield a system of equations for $q_i$ only.

We remark that, in the kinetic terms $A$ and the source $S$, we do not consider the $z$-dependence of the coefficients. In this case one finds simple expressions for the matrices $A$ and $\Gamma$ and for the vector $S$, which can be found in refs.~\cite{Dorsch:2021nje, Dorsch:2021ubz} for any order in the momentum expansion. The system can then be solved using Green's method~\cite{Moore:1995si, Dorsch:2021nje, Dorsch:2021ubz}, yielding
\begin{equation}
    \small
q(z) = v_w\, \chi\cdot \left\{\int_{-\infty}^\infty 
				\dfrac{\partial_{z} m^2}{2T_+^2}
                    e^{-\lambda \gamma_w^{-1} (z-z^\prime)}
				\cdot \theta(\lambda(z-z^\prime))\, dz^\prime\right\}
				\cdot \text{sign}(\lambda)\cdot \chi^{-1} \cdot A^{-1}\cdot S,
\end{equation}
where $\chi$ is a matrix whose columns are the eigenvectors of the matrix $A^{-1}\cdot \Gamma + \sum_i g_i A^{-1}_\text{bg} \cdot \Gamma_i$, and $\lambda$ a diagonal matrix with the corresponding eigenvalues. The background fluctuations can be found immediately from integration of eq.~\eqref{eq:bg} with the boundary condition $q_\text{bg}(z\to -\infty)=0$, i.e. the fluctuations vanish far away ahead of the wall, where the passage of the bubble has not yet been felt.

\begin{figure}
    \centering
    \includegraphics[scale=.46]{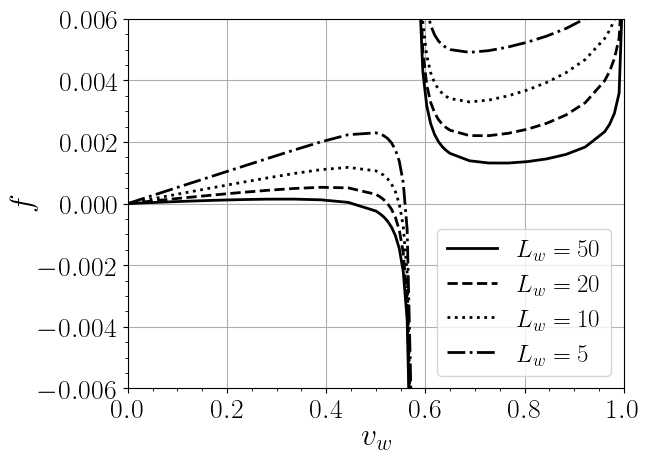}
    \includegraphics[scale=.46]{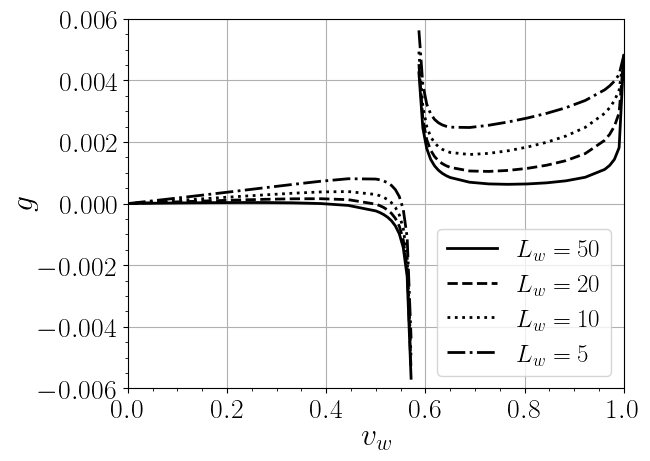}
    \caption{Typical shape of friction terms $f$ and $g$ as functions of wall velocity and varying the wall width $L_w$ (in units of temperature). The transition strength was fixed at $\phi_0/T=2$ for definiteness, and only tops were considered as heavy particles in the fluid. Notice the discontinuous behaviour across the sound speed.}
    \label{fig:friction}
\end{figure}

Plugging the Ansatz~\eqref{eq:delta} and \eqref{eq:linear} into~\eqref{eq:M1M2} one finds
\begin{equation}
    	\begin{pmatrix} f\\ g \end{pmatrix} = \sum_i {g_i}\int \dfrac{\partial_z m_i^2}{2 T_+^2}\, \begin{pmatrix} 1 \\ (2\phi(z) - \phi_0)/T_+ \end{pmatrix} S\cdot (q_i(z)+q_\text{bg}(z))\,dz .
\end{equation}
A typical shape of these functions is depicted in figure~\ref{fig:friction} as a function of the wall velocity $v_w$ for various values of $L_w$. Notice the discontinuous jump across the sound speed, which ultimately stems from the conservation of energy-momentum in the collision terms for the background species. This sort of discontinuity is expected on general grounds from hydrodynamical arguments~\cite{Dorsch:2021ubz}. On the other hand, the singular behaviour (i.e. the fact that the jump is from $-\infty$ to $+\infty$) is likely an artifact of the breakdown of our linearization procedure of the Boltzmann equation. In our results below we will estimate the region where the validity of this linearization is assured.

\section{Hydrodynamical effects and calculation of $T_+$}
\label{sec:hydro}

So far we have only considered the microphysics that allows us to compute the non-equilibrium distribution functions. Now we turn to the question of computing the temperature $T_+$ ahead of the bubble wall. This can be obtained from macroscopic considerations, namely continuity of the total energy-momentum tensor across the bubble interface~\cite{Espinosa:2010hh, Leitao:2010yw}.

The energy-momentum tensor of a scalar field is
\begin{equation}
    T_{\mu\nu}^\phi = \partial_{\mu}\phi \partial_{\nu} \phi - g_{\mu\nu} \left(\frac{1}{2} \partial_\alpha \phi \partial^{\alpha} \phi - V_0(\phi)\right),
\end{equation}
while that of a relativistic fluid in local equilibrium is
\begin{equation}
T_{\mu\nu}^\text{pl} 
=w\, u_{\mu}u_{\nu} - g_{\mu \nu}p_T,
\label{eq:Tmn_pl}
\end{equation}
where $u_\mu$ is the plasma's four-velocity, $w \equiv e+p_T$ is the enthalpy density, $e$ the energy density and $p_T$ the contribution to the pressure due to the plasma particles alone. We are interested in the energy-momentum far ahead and behind the wall, where $\partial_\mu \phi = 0$ (cf. figure~\ref{fig:EWPTshape}), so the total energy-momentum tensor is
\begin{equation}
    T_{\mu\nu} = T_{\mu\nu}^\phi + T_{\mu\nu}^\text{pl} = w\, u_\mu u_\nu - g_{\mu\nu} (p_T - V_0).
    \label{eq:Tmunu}
\end{equation}
Notice that $p_T - V_0 \equiv p$ acts as a total effective pressure, including also the contribution from the free-energy released by the scalar field during the phase transition. Indeed, one can check that $V_0 - p_T = V(T,\phi)$, the thermal effective potential of the scalar field calculated in eq.~\eqref{eq:Veff}~\cite{Konstandin:2014zta}.

Conservation of this total energy-momentum tensor then leads to
\begin{equation}
    \partial_{\mu} T^{\mu\nu} = u^{\nu} \partial_{\mu}(u^{\mu}w) + u^{\mu} w \partial_{\mu}u^{\nu}- \partial^{\nu}p.
\end{equation}

When the bubble has achieved a steady planar configuration, the solution becomes self-similar and can depend only on the dimensionless coordinate $\xi \equiv z/t$. This parameter plays a double role of characterizing both the position along the profile as well as the velocity of the profile velocity at this point, relative to the plasma's rest frame at infinity. Then, in the planar wall case, the four-velocity of the plasma in this frame at a point $\xi$ is $u^\mu = \gamma (1,v(\xi))$, where $\gamma$ is the Lorentz factor associated to the velocity $v(\xi)$. We can also define a perpendicular direction $\overline{u}^\mu = \gamma (v(\xi), 1)$ such that $\overline{u}_\mu u^\mu =0$. Altogether this implies in
\begin{equation}
	u^\mu\partial_\mu = -\frac{\gamma}{t}(\xi-v)\partial_\xi
	\quad\text{and}\quad
	\overline{u}^\mu\partial_\mu = \frac{\gamma}{t}(1-v\xi)\partial_\xi.
\end{equation}
Then, projecting the continuity equation along $u_\mu$ and along the perpendicular direction $\overline{u}_{\mu}$, using $u_\mu u^\mu = 1$ and $u_\nu \partial_\mu u^\nu = \frac12\partial_\mu (u_\nu u^\nu) = 0$, one arrives at
\begin{equation}\begin{split}
	\partial_z T^{00} = 0 &\implies \frac{\partial_\xi e}{w} =\frac{1}{\mu(\xi,v)}\left( 2\frac{v}{\xi (1-v\xi)} + \gamma^2\partial_\xi v\right),\\
	\partial_z T^{0z} = 0 &\implies \frac{\partial_\xi p}{w} 
        = \gamma_w^2 \mu(\xi,v)\partial_\xi v,
    \label{eq:continuity}
\end{split}\end{equation}
where we have defined
\begin{equation}
    \mu(\xi,v)\equiv \frac{\xi-v}{1-v\xi}.
    \label{eq:mu}
\end{equation}

From
\begin{equation}
    w \equiv T\dfrac{\partial p}{\partial T} = T\dfrac{\partial_\xi p}{\partial_\xi T}
\end{equation}
one can rewrite $\partial_z T^{0z}=0$ as an equation for the temperature profile and integrate to obtain
\begin{equation}
    T(\xi) = T(\xi_0)\exp\left(\int_{v(\xi_0)}^{v(\xi)} \gamma^2 \mu(\xi(v), v) dv\right).
    \label{eq:Txi}
\end{equation}
Now, from the definition of the sound speed $c_s^2 = {\partial p}/{\partial e} = {\partial_\xi p}/{\partial_\xi e}$ and equations~\eqref{eq:continuity} one arrives at
\begin{equation}
    2\frac{v}{\xi} = \gamma^2(1-v\xi) \left(\frac{\mu(\xi,v)^2}{c_s^2}- 1\right) \partial_{\xi}v,
    \label{eq:dv_dxi}
\end{equation}
which can be solved for $\xi=\xi(v)$ and plugged back into~\eqref{eq:Txi} for the profile $T(\xi)$.

\subsection{Classification of solutions and discontinuities}

As can be seen from the arguments above, equation~\eqref{eq:dv_dxi} is a central equation for hydrodynamics. In order to understand its solutions, it will be useful to rewrite the continuity of energy-momentum across the bubble wall in yet another form. Let us denote with a subscript $+$ the quantities ahead of the bubble wall, where the phase is still symmetric, and with $-$ the quantities behind the wall, i.e. in the broken phase. In figure~\ref{fig:EWPTshape}, $+$ would correspond to the left and $-$ to the right direction. Then the conservation of eq.~\eqref{eq:Tmunu} becomes
\begin{equation}
    w_+ v_+^2 \gamma_+^2 + p_+ = w_- v_-^2 \gamma_-^2 + p_-
    \quad\text{and}\quad
    w_+v_+\gamma_+^2 = w_-v_-\gamma_-^2,
    \label{eq:continuity2}
\end{equation}
where here $v_\pm$ are velocities measured by an observer in the bubble wall frame. These equations can be combined to yield the following relation between the fluid velocities ahead and behind the wall~\cite{Espinosa:2010hh, Dorsch:2021ubz},
\begin{equation}
    v_+ = \dfrac{1}{1+\alpha}\left( X_+ \pm \sqrt{X_-^2 + \alpha^2 + \frac23 \alpha} \right)
    \quad\text{with}\quad
    X_\pm \equiv \dfrac{3v_-^2 \pm 1}{6 v_-},
    \label{eq:Xm}
\end{equation}
and where
\begin{equation}
    \alpha \equiv \dfrac{ (w_+-3p_+) - (w_- - 3p_-)}{4\rho_\text{rad}}
\end{equation}
defines the ratio of energy released by the transition per energy contained in radiation in the plasma, $\rho_\text{rad} = \pi^2\times 106.75/30$~\cite{Hindmarsh:2020hop, Giese:2020rtr}. 

\begin{figure}
    \centering
    \includegraphics[trim=0 110 0 90, clip, width=.9\linewidth]{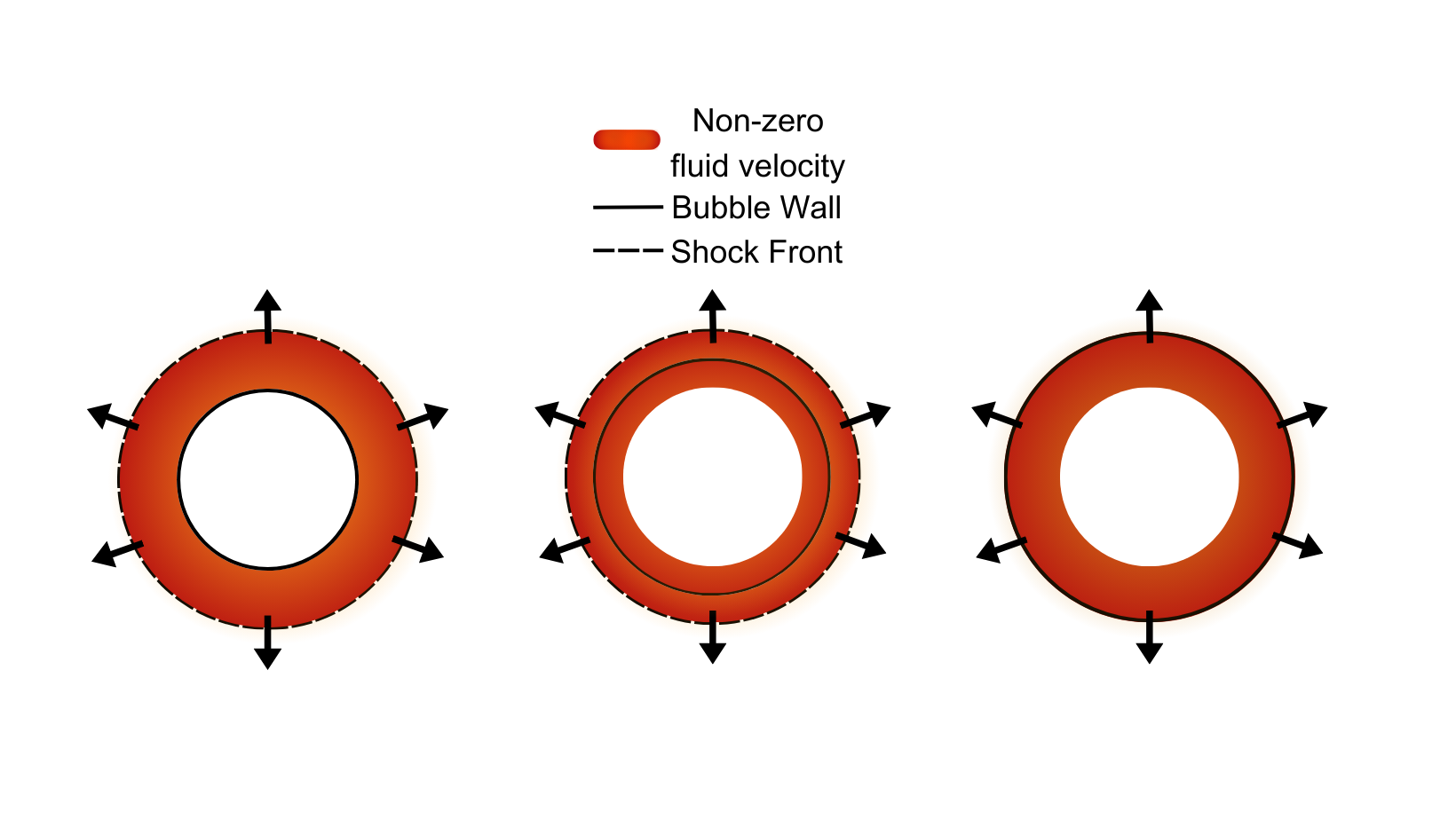}
    \caption{Possible profiles of solutions from the hydrodynamical equation. From left to right: deflagrations, hybrids and detonations.}
    \label{fig:profiles_spherical}
\end{figure}

\begin{figure}[h!]
    \centering
    \includegraphics[scale=.5]{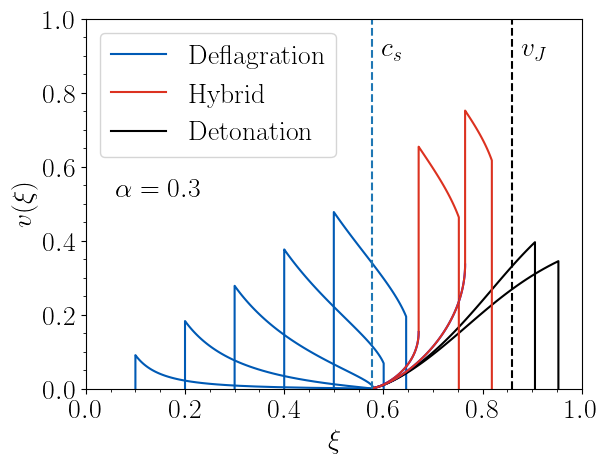}
    \caption{Different velocity profiles $v(\xi)$ for deflagrations, hybrids and detonations. Also shown are the speed of sound $c_s$ and the Jouguet velocity $v_J$. The figure was produced with a fiducial value of $\alpha=0.3$.}
    \label{fig:profiles}
\end{figure}

We are now ready to discuss the solutions of eq.~\eqref{eq:dv_dxi}. It turns out that, depending on the initial condition, the consistent solutions show different patterns of discontinuities, allowing us to classify it into three classes, depicted in figures~\ref{fig:profiles_spherical} and~\ref{fig:profiles}.

One possible class of solutions are detonations. These take place when the wall velocity is larger than the so-called Jouguet velocity\footnote{This is obtained from the Chapman-Jouguet condition that $v_- = c_s$ for detonations~\cite{Steinhardt:1981ct, Espinosa:2010hh, Giese:2020rtr}.},
\begin{equation}
    \xi_w > v_J = \dfrac{1}{\sqrt{3}}\dfrac{ 1 + \sqrt{ 2\alpha + 3\alpha^2}}{1+\alpha}.
\end{equation}
In this case the wall hits the plasma when it is still at rest. This means that the plasma immediately ahead of the bubble has no information about its arrival yet. In particular, the temperature of the plasma immediately ahead of the bubble is equal to the temperature far away from it, i.e. $T_+ = T_n$.

\begin{figure}
    \centering
    \includegraphics{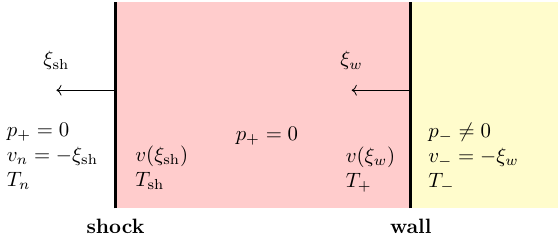}
    \caption{Depiction of the wall and the shock fronts. Behind the wall front (yellow region to the right) the plasma is in the broken phase, $p_-\neq 0$, and at rest, so the plasma velocity measured in the bubble frame is $v_-=-\xi_w$. Ahead of the shock front (white region to the left) the plasma is also at rest, so its velocity in the shock frame is $-\xi_\text{sh}$. The temperature $T_+$ immediately ahead of the wall is found by calculating the jump from $T_n$ to $T_\text{sh}$ across the shock front, and then evolving to the wall front along the shock wave (red region in between) using eq.~\eqref{eq:Txi}.}
    \label{fig:fronts}
\end{figure}

Another class of solutions are deflagrations. In this case the wall propagates at subsonic velocities, $\xi_w\equiv v_w < c_s$, and it is accompanied by a shock wave that precedes the bubble wall and heats up the plasma. This is illustrated in figure~\ref{fig:fronts}. The velocity $\xi_\text{sh}$ of the shock front can be determined in the following manner. In the frame of the plasma\footnote{Also known as ``frame of the bubble center'' in the literature, where the plasma is at rest at infinity/} the plasma is at rest ahead of the shock, and moves with velocity $v(\xi_\text{sh})$ immediately behind it. Transforming to the shock front using the Lorentz law for addition of velocities, eq.~\eqref{eq:mu}, the plasma ahead of the shock moves with velocity $-\xi_\text{sh}$, while behind the shock the velocity is $-\mu(\xi_\text{sh}, v(\xi_\text{sh}))$. Moreover, both ahead and behind the shock the plasma is in the broken phase. Then it can be shown from eq.~\eqref{eq:continuity2} that the velocities satisfy~\cite{Espinosa:2010hh, Leitao:2010yw, Giese:2020rtr}
\begin{equation}
    \xi_\text{sh}\mu(\xi_\text{sh}, v(\xi_\text{sh})) = \frac{1}{3}
	\qquad\text{and}\qquad
	\frac{\xi_\text{sh}}{\mu(\xi_\text{sh}, v(\xi_\text{sh}))} = \frac{3 T_\text{sh}^4 + T_n^4}{3T_n^4 + T_\text{sh}^4}.
	\label{eq:cont_shock}
\end{equation}
The first of these equations can be rephrased as
\begin{equation}
    \xi_\text{sh} = \frac{v(\xi_\text{sh})}{3} + \sqrt{\frac{v(\xi_\text{sh})^2}{9} + \frac{1}{3}}.
    \label{eq:shock}
\end{equation}
Therefore the position of the shock front is found by solving eq.~\eqref{eq:dv_dxi} for $v(\xi)$ and finding the point $\xi$ where $v(\xi)$ satisfies eq.~\eqref{eq:shock} above. Then the second equation in eq.~\eqref{eq:cont_shock} can be used to establish the temperature jump across the shock front,
\begin{equation}
    \frac{T_n}{T_\text{sh}} = \left(\frac{3(1-\xi_\text{sh}^2)}{9\xi_\text{sh}^2-1}\right)^{1/4}.
	\label{eq:Tsh_Tn}
\end{equation}
From eqs.~\eqref{eq:Txi} and \eqref{eq:Tsh_Tn} we have
\begin{equation}
	\frac{T_n}{T_+} = \frac{T_n}{T_\text{sh}}\cdot\frac{T_\text{sh}}{T_+}
	= \left(\frac{3(1-\xi_\text{sh}^2)}{9\xi_\text{sh}^2-1}\right)^{1/4}
		\exp\left(\int_{v(\xi_w)}^{v(\xi_\text{sh})} \gamma^2\mu(\xi(v), v) dv\right),
	\label{eq:Tn_Tp}
\end{equation}
where $\xi(v)$ is found by solving eq.~\eqref{eq:dv_dxi}. For deflagrations the plasma is at rest behind the wall (in the plasma frame), so that in the bubble frame $v_-=-\xi_w$. Thus, given $T_+$ and $v_-$ we can solve the continuity equations~\eqref{eq:continuity2} to find $T_-$ and $v_+$ (the fluid velocity in front of the wall \emph{in the wall frame}). From $v_+$ we can compute the fluid velocity ahead of the wall in the plasma frame using $v(\xi_w) = \mu(\xi_w, v_+)$. Knowing $\xi_w$ and $v(\xi_w)$ we have the initial conditions to solve eq.~\eqref{eq:dv_dxi} for $v(\xi)$ and we can find $v(\xi_\text{sh})$. We can then perform the integration in eq.~\eqref{eq:Tn_Tp} to find $T_n(T_+, v_w)$. For fixed $\xi_w$, we iterate this process until $T_n(T_+, v_w)$ agrees with the nucleation temperature computed according to eq.~\eqref{eq:S_T}. This gives the correct value for $T_+$.

There is yet another kind of solution, called hybrids, which occurs for supersonic walls $\xi_w > c_s$, but differ from Jouguet detonations in that they involve a shock front ahead of the wall, as well as a rarefraction wave behind it. The method for computing $T_+$ in this case is similar to the one for deflagrations delineated above. The difference is that the plasma is no longer at rest behind the wall, but moves at the speed of sound (in the wall frame)~\cite{Espinosa:2010hh, Lewicki:2021pgr}, so one must set $v_- = c_s$ in the continuity equations.

We now have all the tools we need to solve equations~\eqref{eq:M1M2} and find the wall velocity in the model under consideration.

\section{Results and discussions}
\label{sec:results}

One of our main points in the paper is that the behaviour of friction and of the wall velocity will be strongly dependent on the particle content of the theory, even at a qualitative level. In order to show this, we will consider two cases: one where only the top is included as a heavy particle in the effective potential and in the Boltzmann equation (the $W$'s and $Z$'s counting as background, like the other light modes), and another case where tops, $W$'s and $Z$'s are considered heavy.

It has recently been argued in the literature that, once hydrodynamical effects are taken into account, such as the heating of the plasma ahead of the wall, then no non-luminal detonations can be found~\cite{Cline:2021iff}. This is because the pressure against the wall (i.e. $M_1$ in eq.~\eqref{eq:M1M2}) blows up as the wall approaches the speed of sound, and suddenly drops as the Jouguet velocity is reached. So if the wall has enough energy to overcome the Jouguet threshold, the opposite force will not be able to resist its expansion and it will certainly runaway. It is argued that, even if a non-luminal detonation solution exists, a deflagration solution will also exist, and since the wall speed grows from zero up to this stable value, it will stabilize at a deflagration.

\begin{figure}
    \centering
    \includegraphics[scale=.48]{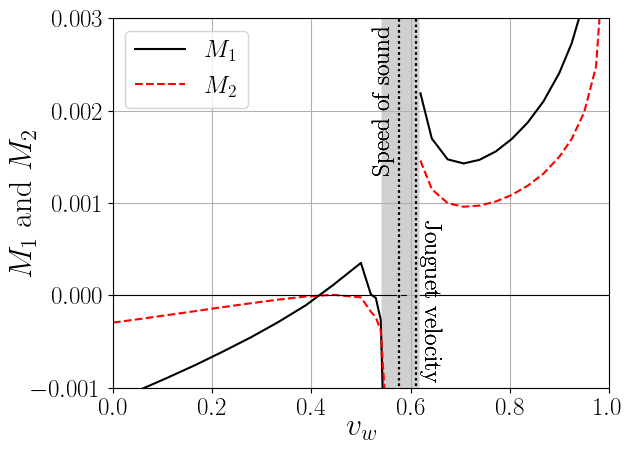}
    \includegraphics[scale=.48]{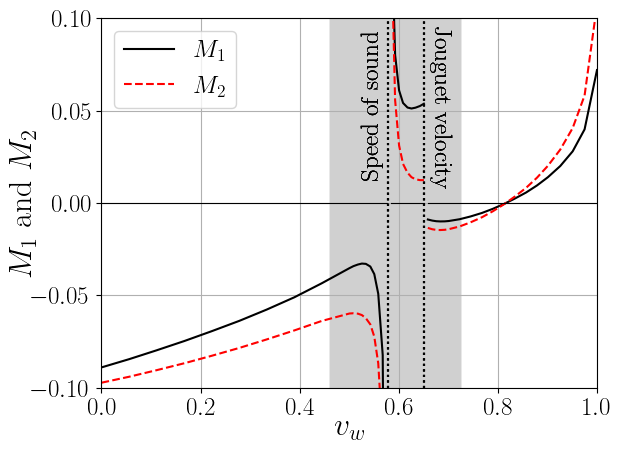}
    \caption{Pressure difference ($M_1$) and ``stretching force'' ($M_2$) across the wall. The left plot corresponds to $M=800$~GeV and $L_wT=18.2354$, and a Jouguet velocity $v_J=0.611$. In this case the solution is a deflagration. The right plot corresponds to $M=630$~GeV and $L_w T=8.2235$, corresponding to a Jouguet velocity $v_J = 0.651$. One notices that the solution is a Jouguet detonation. The behaviour changes drastically across the sound speed due to the shape of the non-equilibrium friction terms, as shown in fig.~\ref{fig:friction} and discussed in depth in ref.~\cite{Dorsch:2021ubz}. Across the Jouguet velocity there is another jump, this time due to hydrodynamical effects, as the expansion is now a detonation and there is no heating of the plasma in front of the wall. In the shaded region we expect the  linearization procedure of the Boltzmann equation to breakdown.}
    \label{fig:M1M2}
\end{figure}

In order to check this statement, we begin by showing in figure~\ref{fig:M1M2} the behaviour of the two ``forces'' $M_1$ and $M_2$ as a function of the wall velocity. Here we consider only tops as heavy particles in the fluid. The left plot corresponds to $M=800$~GeV. In this case one notices that the net pressure does initially increase until both forces are balanced out, then there is a drop across the sound barrier (cf. figure~\ref{fig:friction}). Beyond the Jouguet velocity the non-equilibrium friction dominates, the inner pressure is not enough to push the wall at such high velocities, and there is only one solution corresponding to a deflagration. The right panel, on the other hand, corresponds to $M=630$~GeV. In this case for subsonic walls the friction is never enough to counterbalance the inner pressure, and the only solution is a detonation. Notice that there are two discontinuities affecting this behaviour: the one across the speed of sound (coming from the solution of the Boltzmann equation), and the one across the Jouguet threshold (stemming from the transition of a hybrid to a Jouguet detonation). One could, then, expect to find non-luminal detonations as the only solutions in this case. In shaded grey, we show the region where we expect a breakdown of our linearization procedure of the Boltzmann equation, which can be estimated to occur when~\cite{Dorsch:2021ubz}
\begin{equation}
    \dfrac{\alpha}{X_-^2} \gtrsim 1
    \quad \text{(breakdown of linearization procedure).}
\end{equation}
with $X_-$ defined in eq.~\eqref{eq:Xm}.

\begin{figure}
    \centering
    \includegraphics[scale=.5]{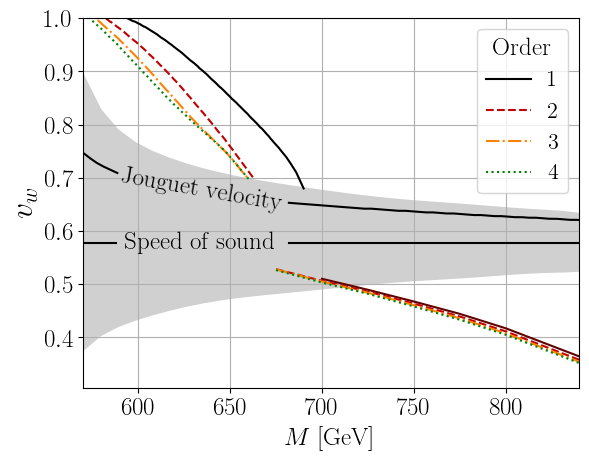}
    \includegraphics[scale=.5]{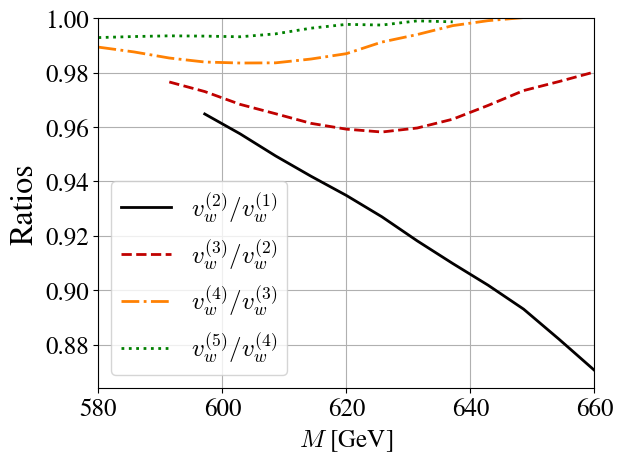}
    \caption{(Left) Solutions for bubble wall velocity as a function of mass scale $M$, including only top quarks as heavy species in the plasma. The shaded region corresponds to the breakdown of the linearization procedure, so solutions in this region cannot be fully trusted. (Right) Ratio of $v_w^{(i+1)}/v_w^{(i)}$, where $v_w^{(i)}$ is the solution at $i$-th order in the fluid expansion. Notice that the solutions quickly converge, as the ratios always decrease and approach unity.}
    \label{fig:vw_M}
\end{figure}

Figure~\ref{fig:vw_M} corroborates the above expectation, showing the wall velocity for a varying cutoff scale $M$ in the case where only tops are considered heavy. The shaded region corresponds to the breakdown of validity of the linearization procedure in the Boltzmann equation.
This means that, outside the shaded region, our procedure is well under control and the results are trustworthy. One notices that there is a range of values for the mass scale $M$ where the transition is strong enough that only non-luminal detonations are possible. Then, for larger values of the cutoff $M$ one approaches the decoupling limit, the transition becomes weaker, and only deflagration solutions exist.

We also show in the right panel of figure~\ref{fig:vw_M} the ratio of velocities, in the detonation regime, computed at two consecutive orders in the expansion of eq.~\eqref{eq:delta}. Notice that the ratio approaches unity as we increase the order of the expansion. This shows that the expansion converges quickly. It is also noteworthy that inclusion of second order effects may reduce the wall velocity by a factor of $\mathcal{O}(10\%)$ compared to a first order calculation, showing the inappropriateness of the perfect fluid \emph{Ansatz}.

\begin{figure}
    \centering
    \includegraphics[scale=.49]{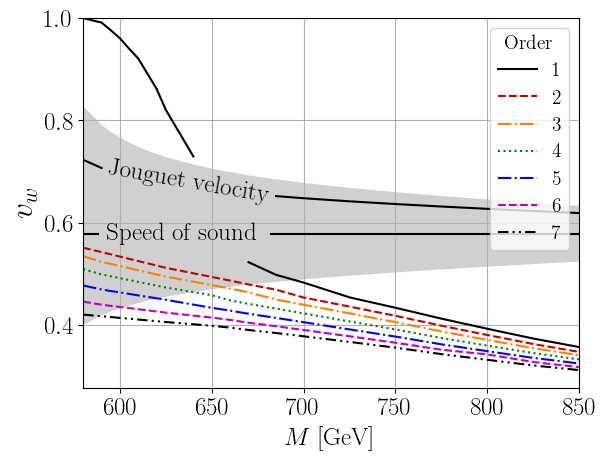}
    \includegraphics[scale=.49]{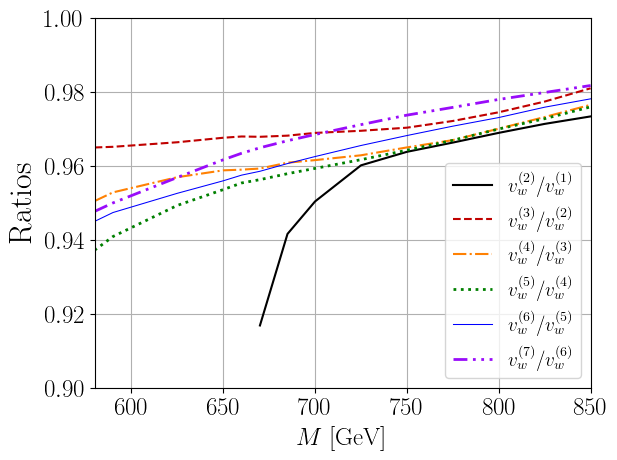}
    \caption{(Left) Solutions for bubble wall velocity as a function of mass scale $M$, including tops, $W$'s and $Z$'s as heavy species. As in figure~\ref{fig:vw_M}, the shaded regions corresponds to the breakdown of the validity of linearizing the Boltzmann equation. (Right) Ratio of solutions of two consecutive orders in the fluid expansion. Notice that the convergence is now much slower, but still granted.}
    \label{fig:vw_M_Wtops}
\end{figure}

Let us now move to the case where the $W$'s and $Z$'s are included as heavy species as well. Here the results are qualitatively different, as can be seen from figure~\ref{fig:vw_M_Wtops}. If we truncate the fluid expansion at first order, i.e. consider a perfect fluid only, one finds similar results to the previous case: deflagrations at large $M$ and detonations for low enough cutoff (strong enough transitions). However, as more orders are included, detonation solutions become untenable. This shows that the inclusion of higher order terms in the fluid expansion can even lead to qualitatively different results, compared to the truncation with three fluctuations only. Moreover, our findings corroborate the recent results of~\cite{Cline:2021iff, Lewicki:2021pgr} that consistent solutions are deflagrations. But notice that the impossibility of non-luminal detonations is not a general statement, but depends on the particle content of the theory.

The right panel of figure~\ref{fig:vw_M_Wtops} illustrates the tendency for convergence of the momentum expansion. We emphasize that this convergence can be expected for any reasonably well behaved function, since eq.~\eqref{eq:delta} is analogous to an expansion in 4-dimensional spherical harmonics~\cite{DeGroot:1980dk}. However, figure~\ref{fig:vw_M_Wtops} shows that, in the presence of gauge bosons, this convergence is much slower than what we observed for tops only. This can be understood from the fact that the expansion parameter is $\mathcal{O}(D/L)$, where $D$ is the diffusion length of the plasma particles and $L$ is the wall width~\cite{Moore:1995si}. For tops $D_\text{top}\simeq 2.9/T$ whereas for gauge bosons $D_\text{W} \simeq 5.5/T$ is almost twice as large~\cite{Moore:1995si, Joyce:1994zt}.

\section{Conclusions}
\label{sec:conclusions}

Determining the velocity of bubble expansion during a first order phase transition is essential for an accurate estimate of the relics stemming from such a process. There have been many recent developments in the literature in this direction, including alternative ways to model non-equilibrium dynamics in the plasma. In this work we study the behaviour of the wall velocity when the friction is evaluated with the so-called ``extended fluid \emph{Ansatz}''. In this approach the non-equilibrium distribution functions have the same functional shape as the equilibrium ones, but while the former depends only on a single power of energy and momentum, the latter instead includes arbitrary powers of a momentum expansion. Since this amounts to an expansion in 4d spherical harmonics, one expects that any reasonably well behaved function could be approximated this way.

We then solve the Boltzmann equation in this approach and compute the fluctuations away from equilibrium, leading to the friction terms. In our solution, we perform a linearization procedure which does not always hold. However, we can establish a criterion for its validity, and can then have an adequate estimate of the reliability of our methods.

Our main result is that the inclusion of higher order terms in the momentum expansion of the fluid \emph{Ansatz} are typically very relevant, and in some cases may even turn a detonation solution into a deflagration. For a Standard Model particle content in the plasma, considering the $W$ and $Z$ bosons and the top quark as heavy species, no detonations are found once we include terms beyond the perfect fluid \emph{Ansatz}, corroborating recent findings in the literature~\cite{Cline:2021iff, Lewicki:2021pgr}. However, this conclusion is heavily dependent on the particle content of the plasma. We illustrate this statement by also analysing a situation where only top quarks are included as heavy species. In this case the overall picture changes and non-luminal detonation solutions become viable solutions. Moreover, in this detonation regime the inclusion of higher order terms in the fluid \emph{Ansatz} is quantitatively important, as the difference to the perfect fluid result may be a factor of $\mathcal{O}(10\%)$ or higher. On a final note, we have checked that the momentum expansion tends to converge, albeit slowly when gauge bosons are included in the picture because of their larger diffusion length compared to the top quarks.

It would be interesting to investigate how the result would be impacted should we include the spatial dependence of the coefficients appearing in the Boltzmann system~\eqref{eq:system}. Similarly, one could also include the spatial dependence of the temperature profile across the bubble wall, which varies from $T_-$ behind the wall to $T_+$ ahead of it. It has recently been argued that this would lead to an equilibrium contribution to friction~\cite{Ai:2021kak}. Recent analyses, based on other models for the non-equilibrium contribution, have reached the conclusion that the equilibrium effects actually dominate over the non-equilibrium term~\cite{Laurent:2022jrs, Ai:2023see}. However, based on the results of~\cite{Ai:2021kak}, we estimate that the equilibrium and non-equilibrium terms should be of comparable size in our approach. We also point out that the results of~\cite{Ai:2021kak} indicate the possibility of non-luminal detonations, but their work does not include the non-equilibrium terms. A full analysis including equilibrium and non-equilibrium contributions in the extended fluid \emph{Ansatz} is already under way, and shall appear in a future publication.

\section*{Acknowledgements}

This work was financed in part by the Coordena\c{c}\~ao de Aperfei\c{c}oamento de Pessoal de N\'ivel Superior - Brasil (CAPES) - Finance Code 001. The authors would also like to thank financial support from the Graduate Program in Physics at UFMG (Brazil) during the preparation of this work.

\end{document}